\documentclass{article}
\usepackage{algorithm2e}

\usepackage{iclr2024_conference,times}
\usepackage{hyperref}
\usepackage{url}
\usepackage{graphicx}
\usepackage{booktabs}
\usepackage{titlesec}

\usepackage{amsmath,amsfonts,bm}

\def\eqref#1{equation~\ref{#1}}

\def\1{\bm{1}}

\DeclareMathAlphabet{\mathsfit}{\encodingdefault}{\sfdefault}{m}{sl}
\SetMathAlphabet{\mathsfit}{bold}{\encodingdefault}{\sfdefault}{bx}{n}

\DeclareMathOperator*{\argmin}{arg\,min}

\makeatletter
\renewcommand\paragraph{
  \@startsection{paragraph}{4}{\z@}{0ex}{-0.5em}{\normalfont\normalsize\bfseries}}
\makeatother

\def\modelname{LiveSpeech 2}

\title{Zero-Shot Text-to-Speech from Continuous Text Streams}

\iclrfinalcopy

\author{Trung Dang\thanks{Correspondence email: trungv.dang@outlook.com}, David Aponte, Dung Tran, Tianyi Chen, Kazuhito Koishida \\
Applied Sciences Group\\
Microsoft Corporation\\
Redmond, WA, USA \\
\texttt{\{trungdang,davidaponte,dung.tran,tianyi.chen,kazukoi\}@microsoft.com} \\
}

\newcommand{\ba}{\boldsymbol{a}}
\newcommand{\bc}{\boldsymbol{c}}

\newcommand{\bz}{\boldsymbol{z}}

\newcommand{\bp}{\boldsymbol{p}}
\newcommand{\bx}{\boldsymbol{x}}

\newcommand{\bh}{\boldsymbol{h}}
\newcommand{\bq}{\boldsymbol{q}}
\newcommand{\by}{\boldsymbol{y}}
\newcommand{\bA}{\boldsymbol{A}}
\newcommand{\bB}{\boldsymbol{B}}
\newcommand{\bC}{\boldsymbol{C}}
\newcommand{\bK}{\boldsymbol{K}}
\newcommand{\bV}{\boldsymbol{V}}

\begin{document}

\maketitle

\begin{abstract}
Existing zero-shot text-to-speech (TTS) systems are typically designed to process complete sentences and are constrained by the maximum duration for which they have been trained. However, in many streaming applications, texts arrive continuously in short chunks, necessitating instant responses from the system. We identify the essential capabilities required for chunk-level streaming and introduce \modelname, a stream-aware model that supports infinitely long speech generation, text-audio stream synchronization, and seamless transitions between short speech chunks. To achieve these, we propose (1) adopting Mamba, a class of sequence modeling distinguished by linear-time decoding, which is augmented by cross-attention mechanisms for conditioning, (2) utilizing rotary positional embeddings in the computation of cross-attention, enabling the model to process an infinite text stream by sliding a window, and (3) decoding with semantic guidance, a technique that aligns speech with the transcript during inference with minimal overhead. Experimental results demonstrate that our models are competitive with state-of-the-art language model-based zero-shot TTS models, while also providing flexibility to support a wide range of streaming scenarios.
\end{abstract}

\section{Introduction}

In recent years, significant advancements have been made in the field of text-to-speech (TTS), evidenced by reports of human parity across both single-speaker \citep{naturalspeech} and zero-shot scenarios \citep{naturalspeech3,valle2}. However, challenges remain in the realm of low-latency streaming zero-shot TTS, where short text chunks are streamed into the model and short audio chunks are streamed out in real-time. Such models are ideal for integration with upstream tasks that emit texts in small chunks such as large language models \cite{achiam2023gpt,team2023gemini} or streaming translation models \citep{barrault2023seamless}. Addressing these challenges could transform live and interactive communication, paving the way for applications such as low-latency speech-to-speech translation, accent conversion, and responsive voice assistants.

While existing models show promising performance in offline inference, they are not suitable or do not support streaming. When it comes to on-device streaming, autoregressive modeling approaches \citep{livespeech1,audiolm,voicecraft} offer an advantage due to the capability of streaming the outputs frame-by-frame. The use of stream-unwary models on streaming inputs involves breaking down the text into short text chunks and condition each generation on previously generated speech, e.g., via prompting. Even when these models are adapted to synthesize from an infinite text stream, several challenges arise in a low-latency scenario: (1) the fixed text condition during inference complicates seamless updates with arriving text chunks, for example, the generation for a text chunk cannot leverage newly arriving context for lookahead; (2) the speech output must catch up with the leading edge of the text stream, requiring the length of generated speech to adapt to the arrival time of text chunks; and (3) the model must process short text chunks while ensuring smooth transitions between their corresponding generated speech segments. In addition to these technical requirements, the speed of inference remains a challenge for inference on the device, since the transformer decoder has to generate a fairly large number of tokens for a single second of audio.

\begin{table}[t]
\caption{Comparisons that highlight the capabilities of our proposed models. Stream-unwary models face numerous challenges when adapting to chunk-level streaming scenarios.}
\label{tab:streaming_capabilities}
\begin{center}
\begin{tabular}{p{0.15\linewidth}p{0.4\linewidth}p{0.4\linewidth}}
\multicolumn{1}{l}{\bf Capability} &\multicolumn{1}{l}{\bf Non-Streaming Models} &\multicolumn{1}{l}{\bf Our Proposed Models} 
\\ \hline
Infinitely long speech streaming & NO support for text streaming in. Texts need to be fixed during generation. Long texts must be segmented into sentences. & allow for a sliding window over long text sequences, retaining only the relevant text in the context for each decoding step. \\ \hline
Text-audio stream synchronization & NO support for duration control. Speech may become out-of-sync with the text stream & allow for generating speech that adjusts to keep pace with the arriving text stream. \\ \hline
Seamless transitions between short speech chunks & NO support for conditioning the current generation on previous outputs, causing non-smooth transitions and style inconsistencies. Even when prompting with previous outputs, a ramp-up time remains essential. Moreover, these models usually only support chunks as long as a full sentence. & allow for smooth transitions between chunks and maintaining consistency in styles to past chunks without ramp-up time. Our model consistantly emits speech frame in near-constant time, independent of the incoming text. It also supports text chunk lengths as short as a single word. \\
\end{tabular}
\end{center}
\end{table}

In this paper, we propose \modelname{ }with additional capabilities to overcome aforementioned challenges. First, we adopt Mamba, a recently developed and highly capable recurrent architecture for sequence modeling, and are the first to demonstrate its competitiveness against transformer-based counterparts at large scale. Mamba maintains an internal state and only takes $O(1)$ complexity to perform a decoding step, thus reducing the inference time compared to transformer-based decoder.  We also reduce the memory length for the reference enrollment speech and transcript by compressing them using a transformer-based speech encoder and a byte pair encoding (BPE) tokenizer, respectively. Second, we propose a cross-attention computation method using rotary positional embeddings, enabling a sliding-window approach on the text. This allows the text condition to be updated at any decoding step and facilitates the generation of content beyond the maximum length for which the model was initially trained. Third, we include semantic tokens together with acoustic tokens in the decoding step outputs and propose inference-time semantic guidance to mitigate the misalignment between text and speech. These improvement enables our models to function reliably with low latency in streaming scenarios, particularly when the upstream task outputs long text in short chunks. Table \ref{tab:streaming_capabilities} highlights the new capabilities and compares ours with non-streaming models. We conduct experiments to demonstrate that our model perform competitively with state-of-the-art non-streaming models in terms of content accuracy, speaker similarity, and general audio quality. Experimental results on the LibriLight and LibriTTS dataset demonstrate that our models achieve superior speaker similarity and overall audio quality while providing flexibility to balance latency and content accuracy in streaming scenarios. Audio samples are available \ificlrfinal
at \url{trungd.github.io/livespeech2}
\else
in supplemental materials.
\fi

\section{Related Works}

Recently, progress in audio and speech generation has focused primarily on the utilization of language models \citep{audiolm,musicgen,valle,valle2,xtts} and diffusion models \citep{naturalspeech,naturalspeech2,naturalspeech3,voicebox,Bai2023ConsistencyTTAAD}, with the debate remaining unsettled. Diffusion models demonstrate their potential by directly generate continuous features without relying on an audio codec, offering high content accuracy and inference speed thanks to the non-autoregressive backbone. On the other hand, language models excel in output streaming \citep{livespeech1}, with recent studies \citep{valle2} claiming to achieve human parity on the LibriTTS and VCTK test sets. Both approaches can generate high-quality outputs in non-streaming mode, where the transcript and enrollment speech are available before the generation process starts. Recent works also explore replacing transformer-based decoders with recurrent architectures \citep{lemerle2024small,halloran2024mamba}, showing comparable performance at smaller scales.

Most research on streamable TTS emphasizes the adoption of fully autoregressive architectures \citep{livespeech1,basetts}, often overlooking the latency caused by sentence formation. When it comes to chunk-level streamable TTS systems,  \cite{dekel2024speak} train a streaming TTS model by distilling from a non-streaming TTS with limited access to future context; however, the architecture does not have a strong zero-shot capability (in fact, it is only demonstrated for a single speaker), and the distillation process only supports one setting for the chunk length and chunk lookahead. Our work demonstrates streaming capabilities similar to those of recent efforts on full-duplex models \citep{ma2024language,moshi,wang2024full}, where speech language models can listen and speak simultaneously; however, while those typically focus on improving interruptibility for conversational models, our goal is to synthesize an existing incoming text stream with minimal latency.

\section{Background}

\subsection{Audio Compression with Residual Vector Quantization (RVQ)}

An audio tokenizer is crucial when using a language model decoder to generate audio. Usually, the audio tokenizer is an audio codec \citep{soundstream,encodec,dac,tfcodec,du2024funcodec,siuzdak2023vocos} with an encoder, a quantizer, and a decoder. The encoder transforms the audio signal into a latent representation of $T$ time steps $\bz_1, \bz_2, ..., \bz_T$, which is recursively quantized by a sequence of quantizers to produce $Q$ codes $\bc_i=[c_{i}^{(1)}, c_{i}^{(2)}, ..., c_{i}^{ (Q)}]$ for each frame feature $\bz_i$. Audio tokens can be generated in the same way as language tokens; however, the amount of tokens poses a challenge of high inference time when being predicted sequentially \citep{audiolm}. MusicGen \citep{musicgen} reduces the number of decoding steps by shifting the codes to predict $Q$ codes in a single step, each of which comes from one in consecutive frames. LiveSpeech \citep{livespeech1} also applies the shifting techniques; however, $Q$ codes are divided into groups that are modeled independently in parallel. Stack-And-Delay \citep{le2024stack} also processes shifted codes in parallel to find a balance between performance and inference speed.

\subsection{Linear-Time Sequence Modeling with Mamba}

Based on Structured State Space Sequence (S4) models \citep{gu2021efficiently}. In general, it involves a continuous system that maps a sequence $x(t)$ to $y(t)$ through a latent state $h(t)$, defined by four parameters $\boldsymbol{\Delta}, \bA, \bB, \bC$, fomulated as $h'(t)=\bA h(t)+\bB x(t), y(t)=\bC h(t)$. After discretizing with zero-order hold: $\overline{A}=\exp(\Delta\bA), \overline{B}=(\Delta\bA)^{-1}\left(\exp\left(\Delta\bA-I\right)\right)\cdot\Delta\bB$, the computation becomes $\bh_t=\overline{\bA}\bh_{t-1}+\overline{\bB}\bx_t, \by_t=\bC\bh_t$, which provides a linear recurrence computation for autoregressive inference. The model can also be computed via global convolution for efficient parallelizable training: $y=x*(C\bB, C\bA\bB, \dots, C\bA^k\bB, \dots)$

Mamba overcomes the linear time-invariance constraint of S4 models, while still maintaining computation efficacy. In particular, the parameters $\boldsymbol{\Delta}, B, C$ are functions of the input, and an efficient hardware-aware implementation is used to replace the global convolution computation.

We adopt Mamba \citep{mamba} as the language modeling component in our model to replace transformers in previous work \citep{valle}. Transformers require attention computation over the past context without any compression, which is computationally inefficient on long sequences, Mamba, on the other hand, summarizes the context into a fixed size state vector via the selection mechanism. We believe that state-space models are the more efficient choice for language modeling of audio tokens since audio tokens are usually long, redundant, and biased towards recency. Rather than storing the entire past generation, a compressed state could provide enough information to ensure smooth frame transition and semantic coherence.

\section{\modelname}

In this section, we present \modelname, our zero-shot TTS model with the streaming capability. The model processes a continuous text stream and outputs codec codes on a frame-by-frame basis. The transcript is delivered in short text chunks, taking into account the timing of arrival. Following the overall architecture of LiveSpeech \cite{livespeech1}, our model contains three main components: a speech encoder that encodes enrollment speech, a text tokenizer and embedder that embed text chunks, and an autoregressive decoder.

The speech encoder is a transformer-based encoder that converts enrollment speech of arbitrary length into a fixed-length sequence of embeddings. The embeddings can remain unchanged or be updated any time during streaming. The primary objective of this encoder is to extract a significantly compressed representation of the entire speech, thereby accelerating the decoding time. The text tokenizer extracts token indices and the text embedder outputs a sequence of token embeddings. We employ the byte pair encoding (BPE) tokenizer from Whisper \citep{whisper} for its extensive coverage and compatibility with upstream Whisper model outputs. We call these tokens word tokens, although some of them do not represent complete words. An end-of-stream token (EOS) is used to signal the end of generation. For the decoder, we employ Mamba \citep{mamba} as an alternative to transformers typically used in related works. In addition to offering competitive performance with a linear-time decoding approach compared to transformers, we posit that speech generation necessitates access not to all tokens in history, but only to a continuously updated state. The decoder integrates information from speech and text embeddings through cross-attention.

To facilitate streaming, we maintain in memory only the current and its neighboring text chunks, updating them continuously as decoding progresses or new chunks arrive. However, the model is trained using fixed transcript with a maximum length, resulting in a disparity between training and inference time. We address the challenges as follows. Section \ref{sec:xa_cond} details our approach to enable dynamic text by assigning positional indices aligned with speech to each word token, and employing rotary positional embeddings when computing cross-attention between speech and text. Section \ref{sec:guidance} introduces a method to prevent misalignment by leveraging monotonic semantic guidance from the transcript. 

\begin{figure}[t]
    \centering
    \includegraphics[width=\textwidth]{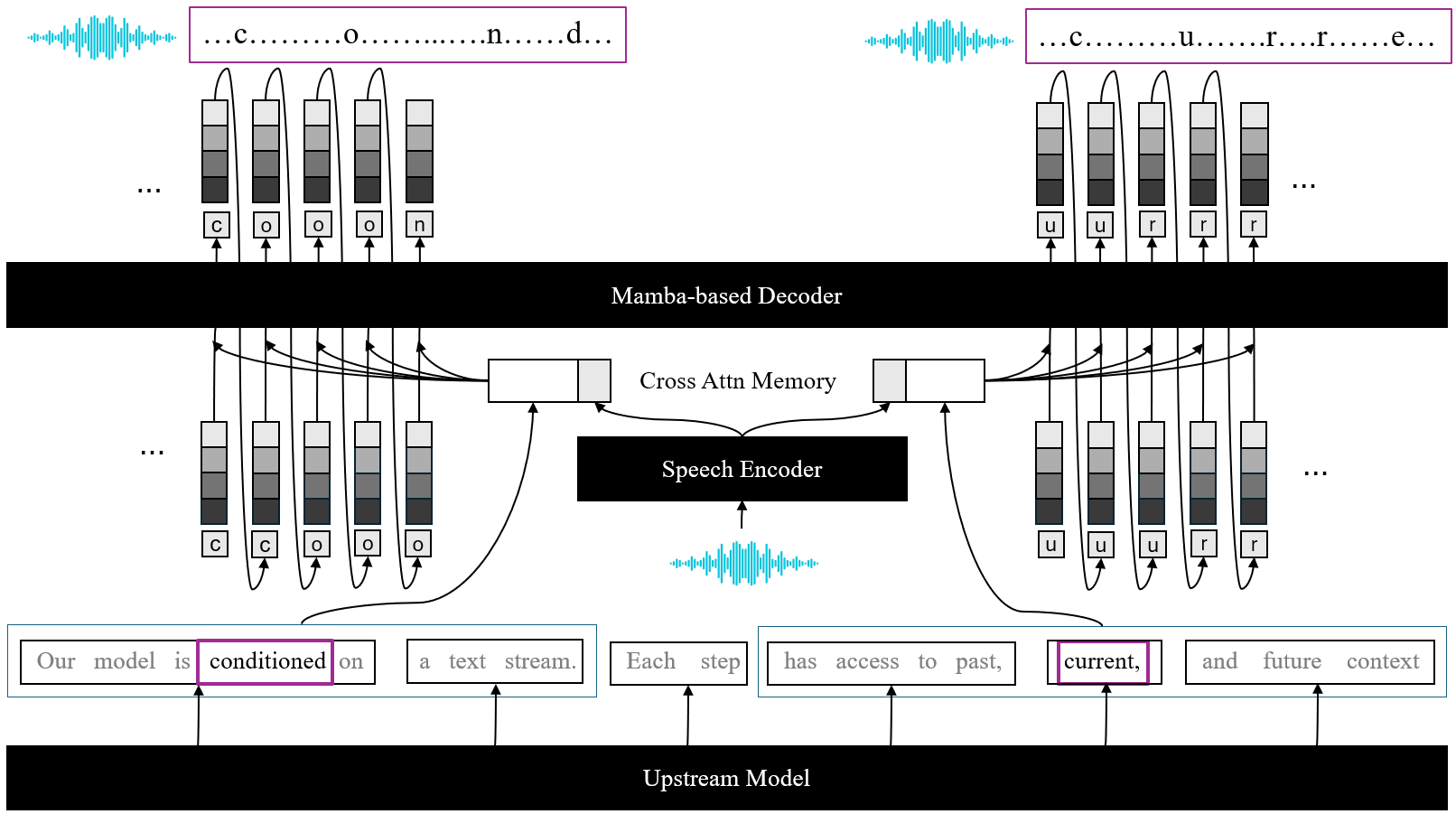}
    \caption{\modelname{ }general architecture. An upstream model generates text continuously in small chunks, while our model synthesizes speech, aiming to keep pace with the most recent chunk. Besides enrollment speech embeddings, each decoding step has access to a section of the text stream, including some past and future chunks.}
    \label{fig:livespeech2}
\end{figure}

\subsection{Text-Speech Cross-Attention with Rotary Positional Embedding}

\label{sec:xa_cond}

Let $S_1, S_2, S_3 \dots$ be chunks from a text stream, where each chunk $S_i$ is a sequence of text tokens $w_i^1,\dots, w_i^{|S_i|}$. Assume that the number of tokens in $S_i$ is bounded by $l_{\text{min}}\le |S_i| \le l_{\text{max}}$. We introduce an additional input $t_i$, which denotes the time between the arrival of chunk $i - 1$ and $i$. To facilitate streaming, the speech for the $i$-th chunk has a duration of approximately $t_i$ in number of frames. Let $\tau_i=\sum_{i'\le i} t_{i'}$, which is the time step at which the chunk $S_i$ starts.

During inference, we aim to add context when new chunks arrive and remove context when it is no longer necessary for generation. However, during training, the context for a single sample is typically fixed for all decoding steps. We adopt a straightforward approach by granting full access to context during training but retaining only certain relevant chunks for each decoding step during inference. We introduce two inference-time hyper-parameters in our system: the maximum number of past chunks included in the cross attention memory, denoted as $n_p$, and the maximum number of future chunks included in the cross attention memory, denoted as $n_f$. In particular, the decoder can attend to $n_p + n_f + 1$ chunks, $(S_{i-n_p},t_{i-n_p}), \dots, (S_i, t_i), \dots, (S_{i+n_f},t_{i+n_f})$, to generate speech for $S_i$ in $t_i$ steps. When $n_f=0$, the system starts generating immediately after a text chunk arrives. When $n_f>0$, the system delays generation until chunk $S_{i+n_f}$ arrives.

\begin{figure}
    \centering
    \includegraphics[width=\textwidth]{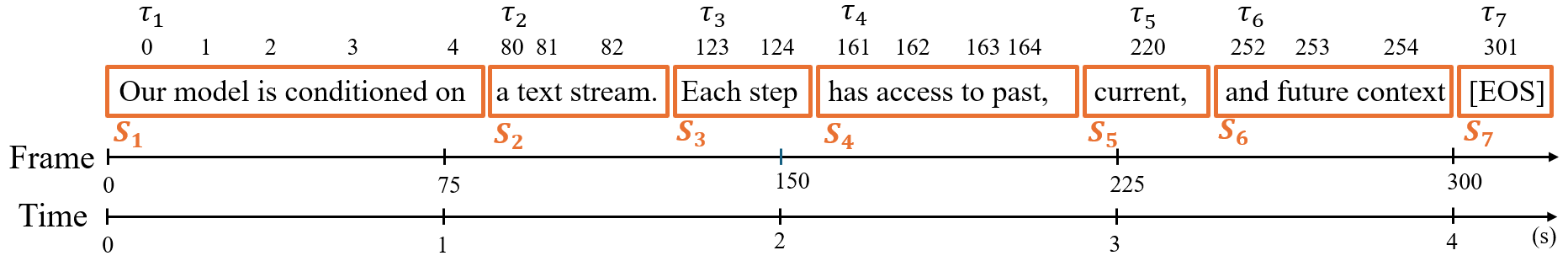}
    \caption{For each word token in a chunk $i$, we assign a position index such that the first word has the index of the frame when the chunk arrives $\tau_i$, and subsequent words have incremental indices $\tau_{i} + 1, \dots$}
    \label{fig:chunking}
\end{figure}

\paragraph*{Positional indices based on arrial time} For each word token embedding, we assign a position index to it: word tokens $w_i^1,w_i^2,\dots, w_i^{|S_i|}$ from chunk $S_i$ are assigned with position indices $\tau_i, \tau_{i+1},\dots \tau_i + |S_i| - 1$. Figure \ref{fig:chunking} illustrates this assignment.

\paragraph*{Cross-attention computation}

Cross-attention scores are computed with the enrollment speech features and the word token embeddings. The enrollment speech features are position-agnostic, while the word token embeddings are coupled with positional indices. Let $c(t)$ be the chunk index at the time step $t$, $p(t)=\min\{|S|,c(t)+n_f\}$ and $f(t)=\max\{0,c(t)+n_f\}$ be the first and the last chunks in the memory for the time step $t$. The attention keys for text at the time step $t$ are expressed as: $\bK^{(\text{txt})}_{t}=\left[\bK_{p(t)};\dots;\bK_{c(t)};\dots;\bK_{f(t)}\right]\in\mathbb{R}^{\left(\sum_{p(t)\le i\le f(t)}|S_i|\right)\times d_k}$. The attention keys for enrollment features are denoted by $\bK_t^{\text{(enr)}}$. Let $\mathcal{T}_t$ be the position indices assigned for each key in $\bK^{(\text{txt})}_t$. With our positional index assignment, $\mathcal{T}_t=\left[\tau_{p(t)},\dots, \tau_{f(t)}+|S_{f(t)}|-1\right]$. For each Mamba layer, let $q_t$ be the layer input at the time step $t$. The cross attention is computed as follows:

\begin{equation}\ba_t = \text{Softmax}\left(\left[\frac{\bq_t\bK^{(\text{enr})T}}{\sqrt{d_k}}\bV^{(\text{enr})};\frac{\text{RoPE}(\bq_t,t)\text{RoPE}(\bK^{(\text{txt})T}_t,\mathcal{T}_t)}{\sqrt{d_k}}\bV^{(\text{txt})}\right]\right),\end{equation}

where $\text{RoPE}(\bK, \mathcal{T})$ rotates the key matrix $\bK$ given indices $\mathcal{T}$. The attention-weighted sum of cross-attention values is integrated with the output of the Mamba layer and given to the subsequent layer.

\subsection{Inference-time Semantic Guidance}

\label{sec:guidance}

Autoregressive TTS decoding suffers from the problem of misalignment, resulting in missing, transpositioning, or repeating content \citep{valle}. To address this issue, we propose providing guidance during inference based on the conditioned transcript. 

\paragraph*{Training} During training, we use time-aligned graphemes as an additional codebook, placed before the first acoustic codebook. Time-aligned graphemes can be obtained from the output of a CTC-based ASR model. Since this output includes plenty of blank tokens, making it sparse in terms of non-blank tokens, we replace each blank token with the first non-blank tokens to the right of the sequence. As an example, grapheme sequence ``abc'' can have a time aligned grapheme sequence of ``aa\textvisiblespace\textvisiblespace\textvisiblespace bbbb\textvisiblespace\textvisiblespace\textvisiblespace\textvisiblespace\textvisiblespace\textvisiblespace cc\textvisiblespace\textvisiblespace'', which will be processed to become ``aaaaabbbbbbbbbbcc\textvisiblespace\textvisiblespace''. Given that there are 75 acoustic tokens per second, while most CTC models generate only 50 tokens per second, we upsample this sequence to align with the number of decoding steps.

\paragraph*{Inference} During inference, we use the previously generated graphemes $G_{t-1}=\left[g_1,\dots,g_{t-1}\right]$ and the transcript to guide the decoding of the next grapheme $g_{t}$. Let $\bp^{(g)}_{t}=\left[p^{(g)}_{t,1},\dots,p^{(g)}_{t,N_g}\right]\in \mathbb{R}^{N_g}$ be the probability distribution predicted for the next grapheme, where $N_g$ is the number of graphemes. We infer a set of guiding tokens $T_{\text{guiding}}$ from the current grapheme sequence and the transcript by determining the prefix of the transcript that matches most closely with $C^{(g)}_{t-1}$. Guiding tokens are either the last token in the prefix (staying) or the next token following the prefix (moving forward). We also infer a set of top-k tokens $T_{\text{top-k}}$ by taking graphemes with highest probability. The next grapheme is sampled from $T_{\text{guiding}}\cup T_{\text{top-k}}$ with a reweighted probability determined by upscaling the probability of guiding graphemes by $(1+\lambda)$ and renormalizing. When $\lambda=0$, \textit{no guidance} is provided. When $\lambda\rightarrow\infty$, we call it \textit{hard guidance} when the next grapheme is only chosen from guiding graphemes. When $0<\lambda\ll\infty$, we call it \textit{soft guidance} where the guiding graphemes are factored in the choice of the next grapheme. In short, we identify a set of graphemes such as if we append one of those to the generated time-align grapheme sequence, this new grapheme sequence has the least CER score to a prefix sequence of the transcript. Hard guidance expects the grapheme sequence to exactly follow the transcript, while soft guidance allows mistakes in the process. Algorithm \ref{alg:guiding_toks} illustrates the sampling process with semantic guidance.

\SetKwComment{Comment}{// }{}
\RestyleAlgo{ruled}
\begin{algorithm}[h]
\caption{Autoregressive decoding with semantic guidance}
\label{alg:guiding_toks}
\KwData{Target transcript $\bar{G}_t=[\bar{g}_1, \bar{g}_2, \dots\bar{g}_{|\bar{G}_t|}]$. Previous decoded graphemes $G_{t-1}=[g_1,g_2,\dots g_{t-1}]$. Softmax probability of the next grapheme $\bp^{(g)}_{t}=\left[p^{(g)}_{t,1},\dots,p^{(g)}_{t,N_g}\right]$. Guiding coefficient $\lambda$. Number of graphemes $N_g$}
\KwResult{Next grapheme $g_t\in [1,...,N_g]$}
$\tilde{G}_{t-1}:=\text{CTCDecode}(G_{t-1})$ \Comment*[r]{remove repetitive/non-char tokens}
$s_{\text{CER}}:=\min_i\left\{\text{CER}(\tilde{G}_{t-1}, \bar{G}_t[\colon i])\right\}$ \Comment*[r]{the best Character Error Rate}

$T_{\text{guiding}}:=\{\}$

\For{$i\in[1,\dots,|\bar{G}_t|]$}{
    \If{$\text{CER}(\tilde{G}_{t-1}, \bar{G}_t[\colon i])=s_{\text{CER}}$}{
        $T_{\text{guiding}}:=T_{\text{guiding}}\cup\{\bar{g}_i, \bar{g}_{i+1}\}$
    }
}
$T_{\text{top-k}}:=\left\{k\mid p_{t,k}^{(g)}\in\text{TopK}\left(\bp_{t}^{(g)}\right)\text{ and }k\not\in T_{\text{guiding}}\right\}$

$\tilde{\bp}:=\bp_t^{(g)}$

$\tilde{\bp}[T_{\text{guiding}}]:=\tilde{\bp}[T_{\text{guiding}}]\times (1+\lambda)$ \Comment*[r]{reweighing probabilities}

$\tilde{\bp}:=\frac{\bp_t^{(g)}}{\sum \bp_t^{(g)}}$ \Comment*[r]{normalize probabilities}

$g_t\sim \text{TopKSampling}(\tilde{\bp}, k)$

\end{algorithm}

While explicitly generating semantic tokens as a transitional ``language'' between the transcript and acoustic tokens has been proposed \citep{audiolm,speakreadprompt}, semantic tokens only serve as the condition to generate acoustic tokens. We take a further step to use the transcript to guide the decoding process in inference time with flexibility.

In this paper, we focus on English as the target language, selecting graphemes as semantic tokens. However, alternative units could be utilized to accommodate a wider spectrum of languages and applications. It is important to choose a unit that allows the transcript to be used to refine candidates in the decoded sequence. Therefore, both graphemes and phonemes are feasible, whereas self-supervised semantic tokens may present certain challenges.

\section{Experiments}

\subsection{Datasets}

For training, we use LibriLight \citep{librilight}, a 60k hour corpus of unlabelled speech for training. We use Whisper v3 (large) \citep{whisper} and wav2vec 2.0 (base) \citep{wav2vec2} to extract the transcript and its word alignment to speech from each training sample. We observe that while Whisper v3 generally produces transcripts with lower error rates and support for punctuation and abbreviations, it occasionally fails catastrophically. Therefore, we use wav2vec 2.0 transcripts and alignments to filter out poor-quality training samples. Specifically, a sample is discarded if the character error rate (CER) between the transcripts from the two models exceeds 0.1 or if the alignments do not match. Each sample is less than 10 seconds in duration, and an enrollment speech of less than 5 seconds from the same speaker is also extracted.

For evaluation, we utilize samples from the test-clean set of LibriTTS. The test set is filtered and divided into two subsets: (1) target speech samples of 3-10 seconds in duration (totaling 2,288 samples, with an average duration of 5.8 seconds), and (2) target speech samples longer than 10 seconds (totaling 1,002 samples, with an average duration of 14.7 seconds).

For both training and evaluation, we simulate a text stream by randomly dividing the transcript into chunks of 2 to 4 word tokens. The alignments from Whisper v3 are used to infer the arrival time of these chunks. The final chunk contains an end-of-stream (EOS) token, with its time set to the duration of the corresponding speech.

\subsection{Model}

We report results using popular baseline models such as: YourTTS \citep{yourtts}, XTTS v2 \citep{xtts}, MetaVoice \citep{metavoice}, SpeechX \citep{speechx}, and LiveSpeech \citep{livespeech1}. All model code and checkpoints are either public (YourTTS, XTTS v2, MetaVoice) or provided by the authors (Speech X, LiveSpeech).

In our model, the speech encoder is a 6-layer 8-head transformer encoder with a hidden dimension of 1024. We prepend 64 empty features to the enrollment speech features to extract a vector sequence of length 64 representing the speech. The Mamba-based decoder consists of 12 layers with a hidden dimension of 1536. The transcript is tokenized with a vocabulary of 51,866 word tokens the same as Whisper \citep{whisper}.
Following LiveSpeech \citep{livespeech1}, the first 6 layers are shared to model all codebooks, and the last 6 layers divide codebooks into 4 groups of 4, 4, 4, 5 codebooks, respectively, which are modeled separately. We also apply a weight based on the codebook prediction performance with $\lambda_{\text{cb}}=0.1$ \citep{livespeech1}. The cross attention has 16 heads with a hidden dimension of 1536. The maximum length for the cross attention memory is 64 + 75, where 64 features belong to enrollment speech and 75 features belong to maximum 75 word tokens in the transcript. Our audio codec, speech encoder, and decoder have 110M, 77M, and 671M parameters, respectively.

\subsection{Training \& Inference}

\paragraph*{Training} We use Encodec to extract acoustic codes at the bit rate of 12kbps or 16 codes/frame and 75 frames/second. Since Encodec is also trained on general audio and music, we train a new decoder specialized in speech on the LibriLight dataset. The model is trained for 2M steps with batch size 32 on 4 A100 GPUs. We employ a learning rate of $5\times 10^{-4}$ with 200k warm up steps \citep{smith2019super}.

\paragraph*{Inference} We perform two modes of inference: offline inference for 3-10s speech and online inference for speech longer than 10s. For offline inference, all past text chunks ($n_p=\infty$), the current, and $n_f=2$ future text chunks are accessible at each decoding step. For online inference, we slide a window over seven chunks, including $n_p=4$ before and $n_f=2$ after the current chunk being generated. Since each chunk has 2-4 words, our system delays 4-8 words after a chunk arrives until its speech can be streamed. If not specified otherwise, we use semantic guidance with $\lambda=1$.

\subsection{Evaluation Metrics}

We evaluate our models in terms of objective and subjective metrics. \paragraph*{Objective Metrics} In terms of content accuracy, we report the Character Error Rate (CER) score with the transcript obtained through the wav2vec2 base model \citep{wav2vec2} and Word Error Rate (WER) score with the transcript obtained via the Whisper v3 model \citep{whisper}. While Whisper v3 is a stronger model that may give us scores closer to human transcripts, wav2vec2 is expected to give more penalty to pronunciation mistakes. In terms of speaker similarity, we report the cosine similarity scores between the generated and the enrollment speaker embeddings using the ECAPA-TDNN model trained on Vox-Celeb \citep{ecapa}. In terms of general speech quality, we report DNSMOS scores \citep{dnsmos}. \paragraph*{Subjective Metrics} We measure Mean Opinion Score in terms of speaker similarity (SMOS) and naturalness (NMOS). For SMOS, we ask each subject to rate the speaker similarity of the enrollment speech and the speech to be evaluated in a scale of 5. For NMOS, we ask each subject to rate the naturalness of the speech in a scale of 5. For each sample, we allow subjects to adjust scores after listening to all audio clips, facilitating relative comparisons between different models. There are 30 short and 30 long samples, each of which is rated by an average of approximately 5 and 3 subjects, respectively.

\subsection{Results}

The results are reported in Table \ref{tab:results}. In terms of CER / WER scores, we are only behind the XTTS v2 baseline, which is trained on a massive amount of internal and public data. Some models have been found to achieve CER and WER scores that surpass even ground-truth samples, indicating that achieving these scores might involve trading off real speech characteristics for improved CER and WER metrics \citep{voicecraft} (e.g., emphasizing clean audio over audio that resembles enrollment speech). Our model achieves the highest SS score, particularly in long speech generation, with a notable improvement of +2.8 points. In terms of subjective metrics, our model outperforms all baselines in both SMOS and NMOS scores, where more significant improvements are also observed for long inputs in the streaming mode.

\begin{table}
  \addtolength{\tabcolsep}{-0.25em}
  \caption{Comparison of our model to the baselines. Each metric is reported with 3-10s / longer than 10s for the target speech. We do not report results of samples longer than 10s for SpeechX, MetaVoice, and LiveSpeech since some samples exceed their maximum context length. For YourTTS and XTTS v2, long transcript is split by Coqui-TTS \citep{coquitts} into smaller ones, which are synthesized separately. Only our model generates speech for all samples in one shot.}
  \label{tab:results}
  \centering
  \begin{tabular}{lccccccc}
    \toprule
    Model & CER & WER & SS & O-MOS & SMOS & NMOS \\
    \midrule
    Ground-truth & 1.6 / 1.4 & 0.6 / 0.7 & 75.3 / 83.9 & 3.9 / 4.0 & 3.8 / 4.0 & 3.7 / 4.0 \\
    Ground-truth (compressed) & 1.7 / 1.5 & 1.0 / 1.2 & 71.1 / 79.3 & 3.9 / 4.0 & 3.5 / 3.9 & 3.5 / 3.9 \\
    \midrule
    YourTTS \citep{yourtts} & 3.8 / 3.3 & 4.3 / 3.6 & 48.6 / 55.2 & 3.8 / 3.9 & 2.6 / 2.3 & 2.5 / 1.9 \\
    XTTS v2 \citep{xtts} & \textbf{1.9} / \textbf{2.2} & \textbf{1.3} / \textbf{1.9} & 60.3 / 64.8 & \textbf{4.0} / \textbf{4.0} & 3.1 / 2.9 & 3.0 / 3.1 \\
    SpeechX \citep{speechx} & 3.8 / ---  & 4.4 / --- & 57.6 / --- & 3.8 / --- & 2.3 / --- & 2.1 / --- \\
    MetaVoice \citep{metavoice} & 4.7 / --- & 4.1 / --- & 56.2 / --- & 3.7 / --- & 2.2 / --- & 2.2 / --- \\
    LiveSpeech \citep{livespeech1} &  3.3 / --- & 6.0 / --- & 59.3 / --- & 3.8 / --- & 2.8 / --- & 2.5 / --- \\
    \midrule
    \modelname{ (ours)} & 2.7 / 3.0 & 3.1 / 4.1 & \textbf{61.7} / \textbf{67.6} & 3.9 / \textbf{4.0} & \textbf{3.4} / \textbf{3.4} & \textbf{3.2} / \textbf{3.3} \\
    \bottomrule
  \end{tabular}
\end{table}

\subsection{Ablation Study \& Analysis}

\paragraph*{The importance of semantic tokens and semantic guidance} We conduct an ablation study when the model does not generate semantic tokens and when they are generated but semantic guidance is not used. Table \ref{tab:results_abl} shows the results. By including semantic tokens in each step, we are able to obtain significant gains in the WER score, especially for long speech where error propagation is more problematic. Semantic guidance also shows considerable effect on the content accuracy, with 53\% improvement in offline scenario and 27\% improvement in online scenario.

\paragraph*{N-time sampling} Existing studies \citep{valle2,naturalspeech2,voicecraft} utilize simple heuristics to select the output from multiple generated outputs; these heuristics range from length-based to metric-based criteria. By incorporating grapheme tokens in our model outputs, transcripts and CER scores of generated speeches become available without the need for an ASR system. Table \ref{tab:results_repeat} illustrates the improvement gains for N-time sampling and compares them with a probability-based criterion, where outputs are selected based on the cumulative probability of the entire sequence of graphemes. Although the probability-based criterion does not guarantee optimal CER scores, it can select the highest in overall probability among those with the same CER scores, thereby resulting in an improved SS score (+1.0). It is important to note that N-time sampling is applicable only for offline inference.

\begin{table}[t]
\parbox{.5\linewidth}{
\addtolength{\tabcolsep}{-0.25em}
  \caption{Ablation study on the necessity of semantic tokens and semantic guidance.}
  \label{tab:results_abl}
  \centering
  \begin{tabular}{lcccccc}
    \toprule
    Model        & 
    WER & SS \\
    \midrule
    With sem guidance  &  
    \textbf{3.1} / \textbf{4.1} & \textbf{61.7} / \textbf{67.6}  \\
    W/o sem guidance  & 
    6.7 / 5.6 & 61.3 / 67.4  \\
    W/o sem tokens & 7.3 / 13.4 & 60.6 / 67.0  \\
    \bottomrule
  \end{tabular}
  }
  \parbox{.5\linewidth}{
  \addtolength{\tabcolsep}{-0.25em}
  \caption{Results when each sample is generated $N$ times and selected based on CER or probability scores.}
  \label{tab:results_repeat}
  \centering
  \begin{tabular}{lcccccc}
    \toprule
    Model        & 
    WER & SS \\
    \midrule
    1-time  &  
    3.1 / 4.1 & 61.7 / 67.6  \\
    2-time (CER based) & 
    2.3 / 3.6 & 61.8 / 67.6  \\
    5-time (CER based) & 
    \textbf{2.0} / \textbf{3.1} & 61.8 / 67.6 \\
    5-time (prob based) & 2.1 / 3.5 & \textbf{61.9} / \textbf{68.6} \\
    \bottomrule
  \end{tabular}
  }
\end{table}

\begin{table}[t]
\parbox{.5\linewidth}{
  \caption{Results for different text chunk minimum ($l_{\text{min}}$ words) and maximum ($l_{\text{max}}$ words) lengths}
  \label{tab:results_chunk_length}
  \centering
  \begin{tabular}{cccccc}
    \midrule
    $l_{\text{min}}$ & $l_{\text{max}}$  & WER & SS \\
    \midrule
    1 & 1 & 40.7 / 73.7 & 58.2 / 65.3 \\
    1 & 3 & 6.8 / 8.1 & 61.6 / 69.5 \\
    2 & 2 & 3.4 / 4.8 & 60.6 / 69.0\\
    2 & 4 & 4.0 / \textbf{3.9} & \textbf{62.2} / \textbf{70.1} \\
    3 & 7 & \textbf{3.6} / 4.5 & 62.1 / 69.8 \\
    \bottomrule
  \end{tabular}
}
\parbox{.5\linewidth}{
  \caption{Results for different text chunk lookback ($n_p$ chunks) and lookahead ($n_f$ chunks)}
  \label{tab:results_num_chunks}
  \centering
  \begin{tabular}{cccccc}
    \toprule
    $n_p$ & $n_f$ & WER & SS \\
    \midrule
    1 & 1 & 23.5 / 15.6 & 61.3 / 68.2    \\
    10 & 1 & 7.5 / 8.5 & 61.5 / 68.9 \\
    2 & 2 &  3.3 / 4.6 & 61.1 / 69.5    \\
    10 & 2 & 3.8 / 3.7 & \textbf{62.3} / 69.8 \\
    10 & 4 & \textbf{3.0} / \textbf{3.3} & 61.3 / \textbf{69.9} \\
    \bottomrule
  \end{tabular}
}
\end{table}

\paragraph*{Effects of the text chunk length} The chunk lengths depend on the upstream task. When only a small local context is required to infer the text (e.g., transcribing), we expect short chunks and lower latency. When the inference of the text requires more global context (e.g., translating), longer chunks are usually needed for better accuracy. For the same transcript, we investigate how different chunking situations affect the quality of the generation. Table \ref{tab:results_chunk_length} shows results for different ranges $[l_{\text{min}},l_{\text{max}}]$. Our model perform poorly in WER score when each chunk has only one word token, hinting that further fine-tuning is required for this extreme scenario. We provide results on streaming aware training in the Appendix \ref{appendix:finetune}, where WER scores are significantly improved even when each chunk has only one word. WER score significantly improves when we increase the range to $[1, 3]$ or $[2, 2]$, and continues to improve as the chunk length increases.

\paragraph*{Effects of the number of text chunks} We investigate the impact of modifying the extent of access to preceding ($n_p$) and succeeding ($n_f$) text chunks on the content fidelity and the audio quality of synthesized speeches. The model exhibits suboptimal performance when constrained to only a single chunk from both preceding and succeeding contexts; however, its efficacy improves with the expansion of access to prior chunks. When the model is allowed to see more of future chunks ($n_f>1$), its performance significantly improves. We also observe an improvement in SS scores when extending the number of past chunks from 2 to 10, suggesting that access to a longer text history enhances certain aspects of voice style.

\section{Conclusion \& Societal Impact}

We introduced \modelname, a zero-shot text-to-speech (TTS) model capable of real-time audio synthesis from continuous textual input. Our model supports real-time applications by continuously streaming short text chunks into the model while producing audio chunks at a constant pace. Given its ability to synthesize speech for any voice, there are concerns regarding possible misuse.

\newpage

\bibliography{iclr2024_conference}
\bibliographystyle{iclr2024_conference}

\newpage

\appendix

\section{Appendix}

\subsection{Data Processing}

From LibriLight, we extract audio clips of up to 30s. For each samples in the batch, we take a random crop of up to 10s starting and ending based on the time-aligned grapheme sequences from pre-decoded CTC model outputs. These outputs are obtained from the large wav2vec2 model pre-trained on the LibriLight and fine-tuned on the LibriSpeech dataset \citep{wav2vec2}. We infer word and chunk level timestamps of whisper transcripts by aligning them with the time-aligned grapheme sequences by the wav2vec model. These timestamp information is used in training and inference to simulate streaming.

\subsection{Test Sets for Additional Study}

For the experiments below, results are reported for small-scale test sets: (1) {[D-off]} Offline Test Set: 94 samples, each lasting 3-10s. (2) {[D-mixed]} Mixed Test Set: no filtering on the target duration. Each chunk contains 3-10 word tokens.

\subsection{Additional Experiments}

We train some variations of our base model:

\begin{itemize}
    \item {[M1]} Base model, as described in the main paper
    \item {[M2]} Numbers of acoustic codes in each group are [2, 3, 4, 8] (versus [4, 4, 4, 5]). With less codes predicted in high-level groups, we expect better quality in high-level (early) codes, which are important to generate better low-level (later) codes.
    \item {[M3]} Different enrollment speech features are utilized in each head. Specifically, the speech encoder outputs 64 features, with each of the 4 heads accessing 16 features. This approach aims to allow each head to learn specific enrollment speech features independently, rather than sharing them.
    \item {[M4]} Only the grapheme decoding has access to the transcript. The other acoustic code decodings can only observe the generated graphemes. There are 1, 4, 4, 8 codebooks in each group, respectively, to allow this condition. This design ensures that acoustic tokens are generated solely based on the grapheme sequence.
    \item {[M5]} We fine-tune our base model [M1] and limit the text context at training time as we do at inference time when using semantic guidance.
    \item {[M6]} We pre-train our base model [M1] while limiting the text copntent at training time as we do at inference time using semantic guidance.
\end{itemize}

We report results in Table \ref{tab:results_additional}. To our surprise, prioritizing early codebook groups [M2] does not improve the CER score, although more capacity is given to predict high-level codes, which are crucial for content accuracy. The [M3] model shows a significant degradation in the CER score but a promising SS score, despite the low CER score. We observe a slight performance decrease in the [M4] model compared to the base model [M1], indicating that access to both the transcript and the predicted grapheme sequence are necessary for the prediction of acoustic codes.

\begin{table}[h]
  \caption{Additional Experiments. Results reported for the D-mixed dataset}
  \label{tab:results_additional}
  \centering
  \begin{tabular}{llcccc}
    ID & Model & CER & WER & SS & DNSMOS \\
    \hline
    {[M1]} & Base & 2.6 & 3.6 & 64.7 & 3.9 \\
    {[M2]} & Prioritizing High-level & 3.4 & 4.1 & 63.3 & 3.9 \\
    {[M3]} & Separating Enrollment Features & 8.1 & 10.6 & 64.9 & 3.8 \\
    {[M4]} & Only Grapheme Seeing Transcript & 3.2 & 4.2 & 63.9 & 3.9 \end{tabular}
\end{table}

\subsection{More Ablation Study \& Analysis}

In this section, we report results on the D-off test set. We provide results for baseline models on this test set in Table \ref{tab:results_abl_offline_baselines}.

\begin{table}[h]
  \caption{Baseline results for the test set used used for offline inference ablation study \& analysis.}
  \label{tab:results_abl_offline_baselines}
  \centering
  \begin{tabular}{lcccc}
    Model & CER & WER & SS & DNSMOS \\
    \hline
    XTTS v2 & 2.0 & 2.9 & 59.0 & 3.95 \\
    YourTTS & 3.5 & 3.7 & 47.2 & 3.82 
  \end{tabular}
\end{table}

\paragraph*{Guidance $\lambda$}

Table \ref{tab:results_guidance} presents the effects of different guidance values. Guidance at any level appears to benefit content accuracy; however, high guidance values may negatively impact the score. Additionally, we observe that high guidance values make it more challenging for the generation to complete full sentences ($\lambda=\infty$ in Table \ref{tab:results_top_k}).

\paragraph*{$k$ in top-k sampling for semantic tokens} Table \ref{tab:results_top_k} shows the effect of limiting the top-k candidates when sampling the semantic tokens with different values of $\lambda$. Overall, there is no clear conclusion on whether sampling benefits more from a larger or smaller number of candidates.

\begin{table}[h]
  \caption{Effect of guidance $\lambda$, $k^{(g)}=5$}
  \label{tab:results_guidance}
  \centering
  \begin{tabular}{lccccc}
    $\lambda$     & CER & WER & SS & DNSMOS \\
    \hline
    0 & 3.1 & 3.5 & 61.0 & 3.85 \\
    1 & 2.4 & 2.9 & 60.7 & 3.85 \\
    2 & 2.6 & 3.2 & 60.9 & 3.85 \\
    3 & 2.5 & 2.9 & 60.8 & 3.85       \\
    4 & 2.6 & 3.1 & 60.9 & 3.85 \\
    5 & 2.7 & 3.0 & 61.2 & 3.85 \\
    6 & 2.9 & 3.0 & 61.1 & 3.85 \\
  \end{tabular}
\end{table}

\begin{table}[h]
  \caption{Effect of $k$ in top-k sampling, dataset [D-off]. $\lambda=0$ means no guidance, $\lambda=\infty$ means hard guidance}
  \label{tab:results_top_k}
  \centering
  \begin{tabular}{llcccc}
    Model & $\lambda$     & CER & WER & SS & DNSMOS \\
    \hline
    $k^{(g)}=5$ & 0 & 2.8 & 3.5 & 60.9 & 3.85 \\
    & 1 & 2.7 & 3.1 & 59.2 & 3.86 \\
    & 2  & 2.5 & 3.1 & 60.9 & 3.86 \\
    & 3  & 2.6 & 3.2 & 61.0 & 3.85       \\
    & $\infty$  & 6.8 & 46.1 & 59.7 & 3.85 \\
    $k^{(g)}=3$ & 0 & 2.8 & 3.7 & 60.7 & 3.85 \\
    & 2 & 2.7 & 2.9 & 61.2 & 3.85 \\
    $k^{(g)}=2$ & 0 & 2.9 & 3.3 & 61.1 & 3.85 \\
     & 10 & 2.5 & 3.2 & 61.1 & 3.84
  \end{tabular}
\end{table}

\subsection{Streaming-Aware Training}

\begin{algorithm}
\caption{Dynamic Cross-Attention Text-Dropout Mask}
\label{alg:attention_window_mask}
\KwIn{kv\_mask, key\_pos\_id, seq\_len, window\_range $(r_1, r_2)$}
\KwOut{mask}

kv\_len $\leftarrow$ Sum(kv\_mask, dim=-1)\;
query\_pos $\leftarrow$ CreateQueryPositions(seq\_len)\;
closest\_text\_pos $\leftarrow \argmin(|$query\_pos - key\_pos\_id$|)$\;

window\_start $\leftarrow$ Floor(Rand() $\cdot$ (closest\_text\_pos - $r_1$)).clamp(min=0)\;
window\_end $\leftarrow$ closest\_text\_pos + $r_2$ + \;
\hspace{1.5em} Floor(Rand() $\cdot$ (kv\_len - closest\_text\_pos - $r_2$).clamp(min=0))\;
window\_end $\leftarrow$ Min(window\_end, kv\_len - 1)\;

key\_pos $\leftarrow$ CreateKeyPositions(kv\_mask.shape)\;
mask $\leftarrow$ (key\_pos $\geq$ window\_start) $\land$ (key\_pos $\leq$ window\_end)\;

\Return mask
\end{algorithm}

\begin{algorithm}
\caption{Applying Cross-Attention Mask and Computing Attention Weights}
\label{alg:apply_mask_and_compute_weights}
\KwIn{pre\_w (results of $Q\*K{^T})$\, kv\_mask, key\_pos\_id, seq\_len, (r1, r2) window\_range}
\KwOut{w (attention weights)}
window\_mask $\leftarrow$ Algorithm \ref{alg:attention_window_mask}(kv\_mask, key\_pos\_id, seq\_len, (r1, r2))\;
kv\_mask $\leftarrow$ CombineMasks(kv\_mask, window\_mask)\;
mask\_values $\leftarrow$ CreateMaskValues(kv\_mask, pre\_w.shape)\;
pre\_w $\leftarrow$ pre\_w + mask\_values\;
w $\leftarrow$ Softmax(pre\_w, dim=-1)\;

\Return w
\end{algorithm}

In the model discussed in the main paper, training is conducted with access to the full context, while online inference is performed with a restricted context, potentially resulting in a mismatch between training and inference conditions. As a result, the model does not perform well in extreme cases when the chunk has only one word, or the model sees one chunk ahead. In this section, we present the results of aligning training with inference by incorporating random context dropout during the training process.

\label{appendix:finetune}

\paragraph*{Fine-tuning for the streaming scenario}
To simulate streaming conditions during training, we implement a dynamic attention masking strategy using Algorithms~\ref{alg:attention_window_mask} and~\ref{alg:apply_mask_and_compute_weights}. This approach modifies the key-value mask to create a visible window centered around the query's closest text position. The algorithms introduce controlled noise by randomly adjusting the start and end points of this visible window, effectively masking out different regions of the text input. This randomized masking simulates the noise and partial information availability characteristic of streaming scenarios.

Our masking strategy employs a ``window range'' parameter that creates a context window around each text token. This window ensures that each query has access to some context from the text, preventing complete information loss.

We conducted ablation studies to assess the effectiveness of this masked fine-tuning strategy. Tables~\ref{tab:ft_results_chunk_length_finetuned} and \ref{tab:ft_results_num_chunks_finetuned} present the results of these studies, where we fine-tuned our base model [M1] using the masking techniques described in Algorithms~\ref{alg:attention_window_mask} and~\ref{alg:apply_mask_and_compute_weights}. These results demonstrate the impact of our dynamic masking approach on the model's performance in streaming scenarios.

\paragraph*{Pre-training for the streaming scenario}
Similar to fine-tuning for the streaming scenario, we conducted ablation studies to assess the effectiveness of Algorithms \ref{alg:attention_window_mask} and \ref{alg:apply_mask_and_compute_weights}.

Tables \ref{tab:ft_results_chunk_length_pretrained} and \ref{tab:ft_results_num_chunks_pretrained} present the results of these studies, where instead of first pre-training [M1] and then fine-tuning to produce [M5], we pretrain [M1] using Algorithms \ref{alg:attention_window_mask} and \ref{alg:apply_mask_and_compute_weights}, producing [M6]. These results demonstrate the impact of our dynamic masking approach on the model's performance in streaming scenarios.

\begin{table}[t]
\small
\setlength{\tabcolsep}{4pt}
\begin{minipage}{.49\linewidth}
  \caption{[M5] Results with different text chunk lengths (online, 44 samples, offline, 94 samples). 
  offline: $n_p$=10,$n_f$=2; online: $n_p$=4, $n_f$=2}
  \label{tab:ft_results_chunk_length_finetuned}
  \centering
  \begin{tabular}{ccccc}
    \toprule
    $l_{\text{min}}$ & $l_{\text{max}}$ & WER & CER & SS \\
    \midrule
    1 & 1 & 4.4 / 5.3 & 3.8 / 4.3 & 57.0 / 64.8 \\
    1 & 3 & \textbf{2.3} / 3.6 & \textbf{2.1} / 3.0 & 57.2 / 64.5 \\
    2 & 2 & 2.9 / \textbf{3.5} & 3.0 / 3.0 & 56.9 / 64.7 \\
    2 & 4 & 2.7 / \textbf{3.5} & 2.7 / \textbf{2.9} & 57.2 / \textbf{65.3} \\
    3 & 7 & 4.7 / 7.4 & 4.1 / 5.8 & \textbf{57.3} / 64.9 \\
    \bottomrule
  \end{tabular}
\end{minipage}
\hfill
\begin{minipage}{.49\linewidth}
  \caption{[M5] Results with different numbers of text chunks}
  \label{tab:ft_results_num_chunks_finetuned}
  \centering
  \begin{tabular}{ccccc}
    \toprule
    $n_p$ & $n_f$ & WER & CER & SS \\
    \midrule
    1 & 1 & 3.5 / 4.6 & 3.2 / 3.1 & 56.0 / 64.9 \\
    10 & 1 & 3.1 / 4.7 & 3.1 / 3.6 & 56.4 / 64.6 \\
    2 & 2 & \textbf{2.7} / 4.5 & \textbf{2.6} / 3.8 & 56.5 / 64.5 \\
    10 & 2 & \textbf{2.7} / 4.9 & 2.7 / 3.5 & \textbf{57.2} / \textbf{65.1} \\
    10 & 4 & 2.9 / \textbf{3.3} & 2.7 / \textbf{2.7} & 56.5 / 64.4 \\
    \bottomrule
  \end{tabular}
\end{minipage}
\end{table}

\begin{table}[t]
\small
\setlength{\tabcolsep}{4pt}
\begin{minipage}{.49\linewidth}
  \caption{[M6] Results with different text chunk lengths (offline, 94 samples, online, 44 samples). 
  offline: $n_p$=10,$n_f$=2; online: $n_p$=4, $n_f$=2}
  \label{tab:ft_results_chunk_length_pretrained}
  \centering
  \begin{tabular}{ccccc}
    \toprule
    $l_{\text{min}}$ & $l_{\text{max}}$ & WER & CER & SS \\
    \midrule
    1 & 1 & 5.4 / 6.8 & 4.9 / 5.1 & 59.9 / 68.4 \\
    1 & 3 & 4.0 / 4.7 & 3.2 / 3.9 & 60.3 / 67.8 \\
    2 & 2 & \textbf{3.3} / \textbf{3.6} & \textbf{2.9} / \textbf{2.5} & \textbf{61.1} / \textbf{69.4} \\
    2 & 4 & 3.3 / 5.1 & 2.9 / 3.8 & 61.1 / 68.9 \\
    3 & 7 & 4.4 / 8.4 & 3.4 / 5.9 & 60.7 / 68.4 \\
    \bottomrule
  \end{tabular}
\end{minipage}
\hfill
\begin{minipage}{.49\linewidth}
  \caption{[M6] Results with different numbers of text chunks (offline, online)}
  \label{tab:ft_results_num_chunks_pretrained}
  \centering
  \begin{tabular}{ccccc}
    \toprule
    $n_p$ & $n_f$ & WER & CER & SS \\
    \midrule
    1 & 1 & 3.3 / 4.98 & 3.5 / \textbf{3.5} & 60.6 / 68.0 \\
    10 & 1 & 4.0 / 4.98 & 3.7 / 3.98 & 60.6 / 67.6 \\
    2 & 2 & \textbf{3.2} / 5.1 & \textbf{2.7} / 3.8 & \textbf{60.6} / 69.0 \\
    10 & 2 & 3.4 / 5.5 & 3.2 / 4.2 & 60.33 / 69.0 \\
    10 & 4 & 3.3 / 5.4 & 2.8 / 3.8 & 60.5 / \textbf{69.2} \\
    \bottomrule
  \end{tabular}
\end{minipage}
\end{table}

\subsection{Areas for Improvement}

\paragraph*{Separate grapheme token prediction from acoustic token prediction} Prediction of the grapheme sequence given the word sequence should not be so challenging; however, in many cases, we observe that the model does not produce the correct grapheme sequence. We hypothesize that error in acoustic codes may affect the accuracy of grapheme prediction, which in turns adversely affect the acoustic codes. By making graphemes not depending on previously decoded acoustic codes, we can potentially improve the accuracy of predicting them. Similarly, high-level codes can be made independent of low-level codes to avoid being affected by their errors.

\paragraph*{Hard guidance may work better for transformer} Hard guidance avoids sampling the wrong candidate for the next grapheme; however, in some cases, the probability for these guiding tokens are low. In state space models, choosing a low probability candidate may hurt more than in transformers, since we can only ``force'' the input but not the internal state.

\subsection{Attention Visualization}

\begin{figure}
    \centering
    \includegraphics[width=1.0\linewidth]{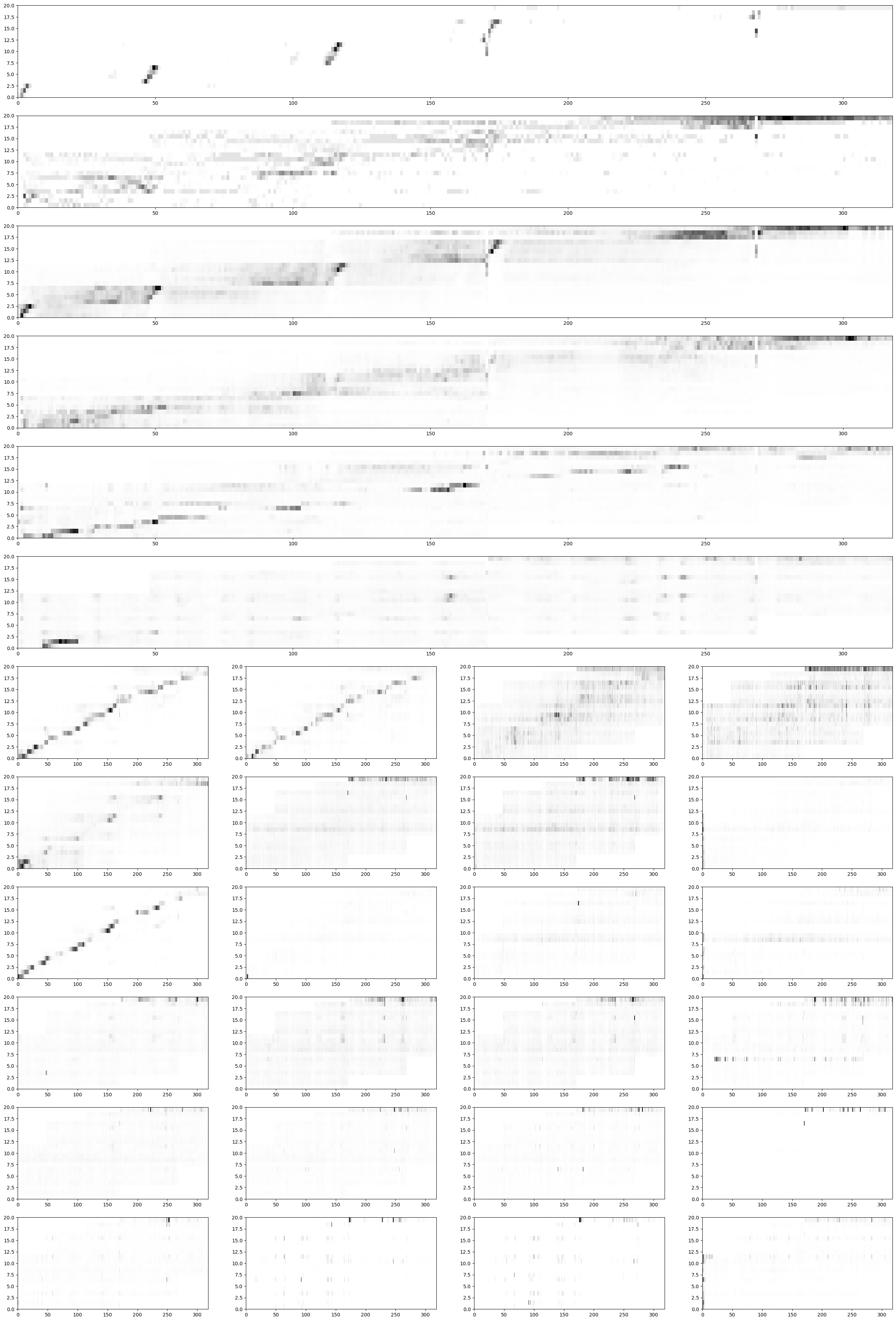}
    \caption{Cross-attention visualization for ``There is even / a white row of / beehives in the / orchard under the walnut / trees". There are 12 rows for each mamba layer, in which each in the first 6 rows has only one head and each in the last 6 rows has four heads, each predicting 4, 4, 4, 5 codes (total 1 grapheme token + 16 acoustic codes) in a frame, respectively. In each plot, the x-axis represents 318 audio frames generated and the y-axis represents 21 word tokens in the transcript.}
    \label{fig:visualize_short}
\end{figure}

\begin{figure}
    \centering
    \includegraphics[width=1.0\linewidth]{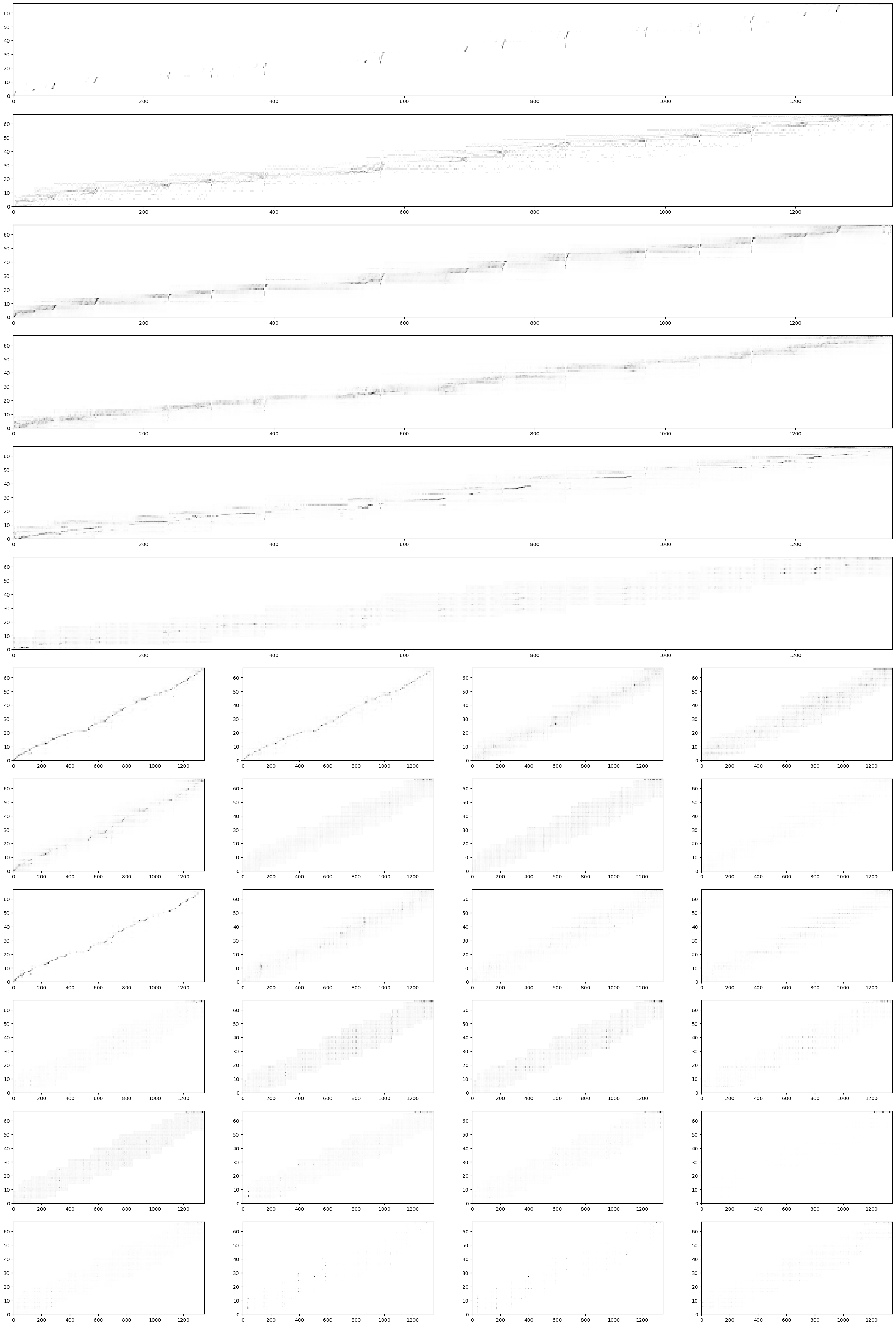}
    \caption{Cross-attention visualization for ``When you go / out of /the house into the / flower garden, there you / feel again the / order and fine / arrangement manifest all over / the great / farm, in the fencing / and hedging, / in the windbreaks and / sheds, in the symmetrical / pasture ponds / planted with scrub / willows to give shade / to the cattle / in fly-time.''}
    \label{fig:visualize_long}
\end{figure}

We provide cross-attention visualization for a short speech (Figure \ref{fig:visualize_short}) and a long speech (Figure \ref{fig:visualize_long}). The first six rows are attention visualization for the first six shared layers. Each of the last six rows presents four plots for codebook groups. For the model reported in the main paper, these four groups contain 4, 4, 4, 5 codebooks, respectively. In the first layer, it is observed that the speech frames tend to align with the word tokens that share similar positional indices. Specifically, the initial frame aligns with the first word in a chunk since they have the same positional indices. However, because a speech chunk contains significantly more frames than there are word tokens in a text chunk, most alignment occurs within the first few frames of the speech chunk. As we go deeper into the model layers, the alignment extends across the entire speech chunk. The alignment is also observed to be less noisy in the first group of codebooks, suggesting that word tokens hold greater significance for this group compared to others.

\subsection{High CER samples}

Table \ref{tab:sample_short} and \ref{tab:sample_long} list samples with highest CER scores from the subjective evaluation set for short and long utterances. It is observed that when the model occasionally makes pronunciation mistakes, especially on hard words, it can mostly avoid problems caused by misalignment such as word repetition or early finishing / hallucinating.

\begin{table}[h]
    \centering
    \begin{tabular}{p{0.4\linewidth}p{0.4\linewidth}p{0.05\linewidth}}
\toprule
\multicolumn{1}{l}{\bf Generated} &\multicolumn{1}{l}{\bf Reference} &\multicolumn{1}{l}{\bf CER} 
\\ \midrule
    he keeps \textbf{that the shot} not \textbf{command mats} first rate \textbf{and} lord does & he keeps the thou shalt not commandments first rate hen lord does & 18.5 \\ \midrule
    but mormonism died \textbf{but} every \textbf{taint} of grief served but to unite the people & but mormonism died not every added pang of grief served but to unite the people & 13.9 \\ \midrule
    was \textbf{it} the bible \textbf{oshed} whispered bill harmon & was that the bible osh whispered bill harmon & 11.4 \\ \midrule
    all \textbf{angelleaks} folks are baking for it and all \textbf{emittes} twenty cousins & all angeliques folks are baking for it and all amitys twenty cousins & 10.3 \\ \midrule
    im afraid \textbf{as when} quite answer the purpose said his \textbf{mamma} smiling especially the last yet we must think of something & im afraid those wouldnt quite answer the purpose said his mama smiling especially the last yet we must think of something & 8.3 \\ \midrule
    the \textbf{beggas} plea the politicians \textbf{sceptre} and the drummes \textbf{is} ablest assistant & the beggars plea the politicians scepter and the drummers ablest assistant & 8.1 \\ \midrule
    every word fell distinctly in perfect harmony and eloquence upon \textbf{lewis exors} ears & every word fell distinctly in perfect harmony and eloquence upon louis xivs ears & 6.3 \\
\bottomrule
    \end{tabular}
    \caption{Samples with highest CER scores from 58 short samples}
    \label{tab:sample_short}
\end{table}

\begin{table}[h]
    \centering
    \begin{tabular}{p{0.4\linewidth}p{0.4\linewidth}p{0.05\linewidth}}
\toprule
\multicolumn{1}{l}{\bf Generated} &\multicolumn{1}{l}{\bf Reference} &\multicolumn{1}{l}{\bf CER} 
\\ \midrule
        the invention is in universal use to day alike for direct and for alternating current and as well in the equipment of large buildings as in the \textbf{inocion} distribution \textbf{[]} of the most extensive central station \textbf{metaress} & the invention is in universal use today alike for direct and for alternating current and as well in the equipment of large buildings as in the distribution system of the most extensive central station networks & 10.0 \\ \midrule
        yet sometimes in the pauses of his work the young man frowned and looked \textbf{up} the ground with an \textbf{intent intent} which suggested that even twenty one might have its problems \textbf{it} & yet sometimes in the pauses of his work the young man frowned and looked at the ground with an intentness which suggested that even twentyone might have its problems & 6.7\\ \midrule
        he hoped there would be stew for dinner turnips and carrots and bruised potatoes and fat mutton pieces to be \textbf{laddled} out in thick peppered \textbf{flurfished} sauce \textbf{suffered} into you his belly counselled him & he hoped there would be stew for dinner turnips and carrots and bruised potatoes and fat mutton pieces to be ladled out in thick peppered flourfattened sauce stuff it into you his belly counseled him & 6.5 \\ \midrule
        i sent the hurons he said speaking to the \textbf{mo ends} yonder \textbf{yonders} open sky through the tree tops and we are getting too nigh their encampment sagamore you will take the hillside to the right \textbf{huncas} will bend along the \textbf{brck} to the left while i will try the trail & i sent the hurons he said speaking to the mohicans yonder is open sky through the treetops and we are getting too nigh their encampment sagamore you will take the hillside to the right hunkus will bend along the brook to the left while i will try the trail & 6.3 \\
        \bottomrule
    \end{tabular}
    \caption{Samples with highest CER scores from 30 samples}
    \label{tab:sample_long}
\end{table}

\end{document}